\documentclass[aps,pra,groupedaddress,reprint]{revtex4-1}
\usepackage{graphicx}
\usepackage{amsmath, epsfig}
\usepackage{amsfonts}
\usepackage{amssymb}
\usepackage{epstopdf}
\usepackage{setspace}
\usepackage{verbatim}
\usepackage{color}
\usepackage{wasysym}
\usepackage[dvipsnames]{xcolor}
\usepackage{array}

\begin{document}
\title{Microwave-optical two photon excitation of Rydberg states}

\author{D. A. Tate}
\affiliation{Department of Physics and Astronomy, Colby College, Waterville, ME 04901-8858, USA}
\author{T. F. Gallagher}
\email{tfg@virginia.edu}
\affiliation{Department of Physics, University of Virginia, Charlottesville, VA 22904-0714, USA}
\date{\today}

\begin{abstract}

We report efficient microwave-optical two photon excitation of Rb Rydberg atoms in a magneto optical trap. This approach allows the excitation of normally inaccessible states and provides a path toward excitation of high angular momentum states. The efficiency stems from the elimination of the Doppler width, the use of a narrow band pulsed laser, and the enormous electric dipole matrix element connecting the intermediate and final states of the transition. The excitation is efficient in spite of the low optical and microwave powers, of order 1 kW and 1 mW, respectively. This is an application of the large dipole coupling strengths between Rydberg states to achieve two photon excitation of Rydberg atoms.

\end{abstract}

\maketitle

 \section{Introduction}

Reaching optically inaccessible Rydberg states of high principal quantum number $n$ is of interest for several reasons. First, a Rydberg atom in a state of high angular momentum $\ell$ state provides a benign probe of the ion core \cite{free76}. For example, the polarizability of the ion core can be determined with high accuracy from the quantum defects of the non penetrating high $\ell$ Rydberg states \cite{mayer33,edlen64,free76}. Second, excitation of atoms to the nearly degenerate high $\ell$ Rydberg states in a dense cold gas of ground state Rydberg atoms leads to the production of ``trilobite'' molecules, composed of one ground state atom and one Rydberg atom, the latter of which is in a superposition of high $\ell$ states \cite{gree00,bend09,tall12}. Since the Rb $nf$ states have quantum defect $\delta_f=0.015$, their excitation, as described here, appears to be the most straightforward route to trilobite molecules \cite{gree00,tall12}.

Finally, it appears possible to cool and trap Rydberg atoms of alkaline earth atoms by using the strong optical resonance transitions of the ionic core. Laser cooling using the ion resonance transition has been proposed as a way to reach the strongly coupled regime in ultra-cold plasmas made from alkaline earth atoms \cite{kill03}, and has been suggested as a means to trap Rydberg alkaline earth atoms \cite{chei18}. For example, bound Sr $5sn\ell$ Rydberg states can in principle be trapped using the strong $5sn\ell \rightarrow 5pn\ell$ transition, which is the resonance line of Sr$^+$ with a spectator $n\ell$ electron \cite{cooke78}. If the Rydberg electron is in a state of low $\ell$, the resulting $5pn\ell$ state decays rapidly by autoionization, and trapping is impossible. However, the autoionization rates of $5pn\ell$ states decrease rapidly with $\ell$, and for high $\ell \sim n$, the $5pn\ell$ atoms decay radiatively, and trapping should be feasible \cite{jacob79,cooke78,jones88}.

High $\ell$ Rydberg states of low $m_\ell$, where $m_\ell$ is the azimuthal orbital angular momentum quantum number, have been produced by Stark switching, in which laser excitation occurs in the presence of an electric field which is then reduced adiabatically to zero \cite{free76}. This technique has been realized by several groups \cite{cooke78,jones88,pruv91,eich92}. Typically Rydberg atoms of $10<n<20$ are excited in a field of order 1 kV/cm, and the field is reduced to zero on a time scale of several microseconds. The primary limitation of the technique is that the electric field must be large enough to mix oscillator strength from an optically accessible low $\ell$ state into the desired Stark state and produce large enough energy splittings that the Stark states can be resolved in laser excitation. In the experiments cited above the laser linewidths were in all cases greater than 2 GHz \cite{pruv91}, which dictated the use of kV/cm fields to allow the Stark levels to be energetically resolved. It is difficult to reduce a 1 kV/cm electric field to produce a stable zero field environment when the field is turned off. A narrower linewidth laser removes energy resolution as a limitation, but the requirement of admixing low $\ell$ oscillator strength remains. If a higher angular momentum state with a small quantum defect is populated by the laser excitation, much smaller Stark switching fields can be used to produce higher $\ell$ states with microwaves  \cite{shum07}.

It is also of considerable interest to produce circular states of cold atoms, those in which $|m_\ell| = \ell = n-1$. The most common approach has been laser excitation followed by microwave adiabatic passage to the circular state \cite{hulet83}. Recently a Rabi oscillation technique has been shown to allow more rapid preparation of the circular state \cite{sign17}.  Both of these approaches depend upon starting from a state with a small quantum defect, such as a Na $nd$ or Rb $nf$ state. The optically accessible Rb $ns$ and $nd$ states are not usable, although the Rb $ns$ states could be used in a variant of Stark switching using a circularly polarized microwave field \cite{cheng94}.

\begin{figure}
\centerline{\resizebox{0.35\textwidth}{!}{\includegraphics{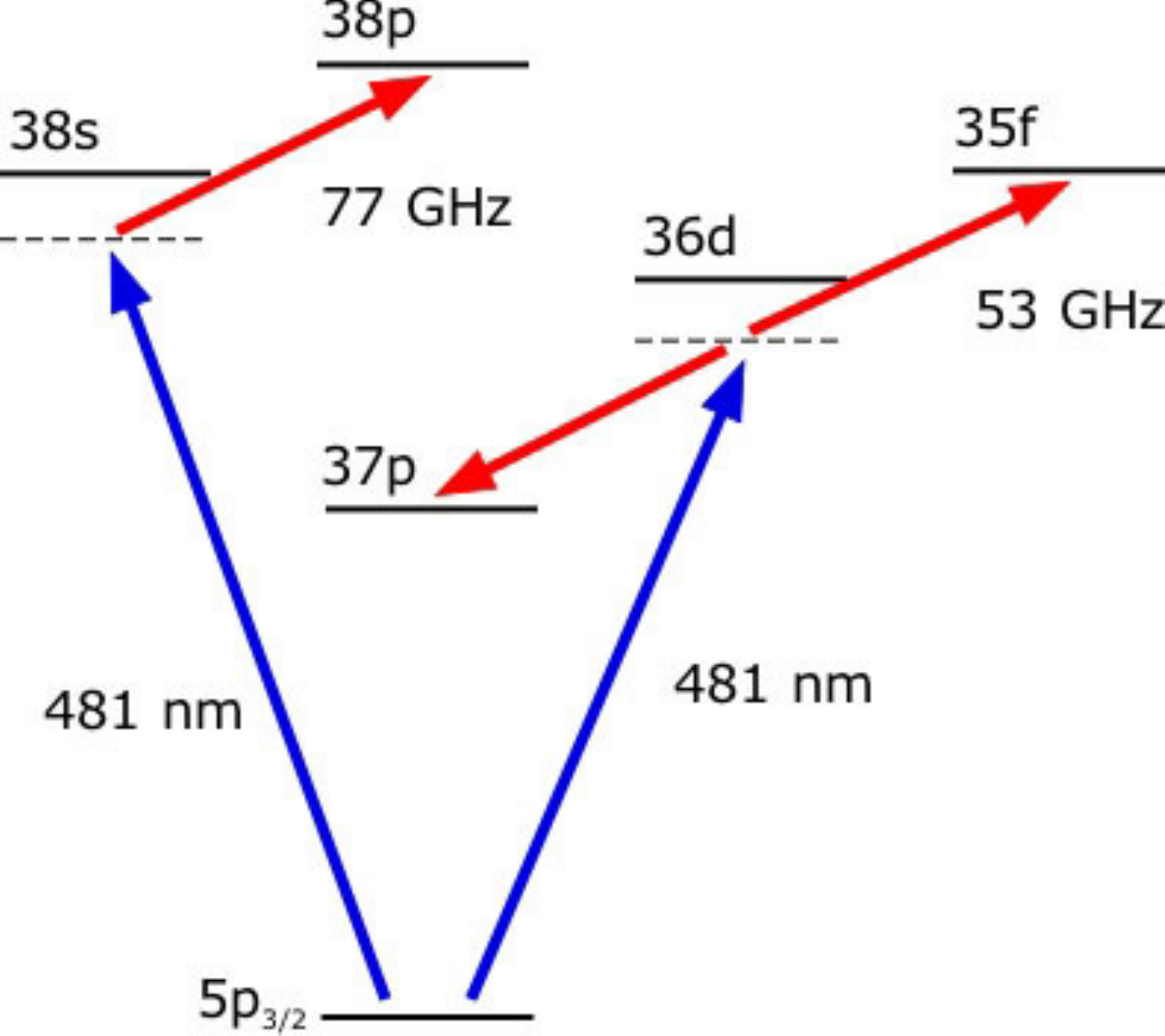}}}
\caption{(color online).
Schematic of the Rb energy levels and transitions used in this work.
The blue arrows show the 481 nm laser contributions and the red arrows the microwave contributions.}
\label{termdiag}
\end{figure}

Here we describe two photon microwave-optical excitation of cold Rb Rydberg atoms in a magneto optical trap (MOT). Specifically, using a 480 nm optical pulse and a 50 - 80 GHz microwave pulse we have driven the Rb $5p_{3/2} \rightarrow 35f$, $5p_{3/2} \rightarrow 37p_j$ and $5p_{3/2} \rightarrow 38p_j$ transitions, as shown in Fig. \ref{termdiag}. The Rb $nf$ states have a quantum defect of $\delta_f=0.015$, so it is far easier to reach the high $\ell$ states from the $nf$ states than from the optically accessible $nd$ states, which have a quantum defect of $\delta_d=1.35$. The excitation is efficient in spite of the use of modest optical and microwave powers, $\sim1$ kW and $\sim1$ mW, respectively. Several factors allow efficient two photon excitation with such modest sources. First, having cold atoms in a MOT eliminates Doppler broadening. Second, the relatively narrow bandwidth, 150 MHz (FWHM), of the 480 nm pulse allows us to take advantage of the small Doppler broadening and use small detunings, $\sim$1 GHz, while maintaining good state selectivity in the excitation. Finally, the transition from the near resonant intermediate state to the final state has an enormous electric dipole matrix element, of order $n^2$ atomic units \cite{gall94}. This is a use of the large dipole coupling strengths between Rydberg states in their excitation using lasers. A variant of this approach for Rydberg electromagnetically induced transparency has recently been demonstrated  \cite{vogt18}.

In the sections which follow we describe our experimental approach, present the spectra we have obtained, and suggest possible future improvements.

\section{Experimental approach}

Our vapor cell MOT apparatus has been described previously \cite{li03}. Cold $^{85}$Rb atoms at a density of $\sim10^9$ cm$^{-3}$ are held in the $\sim$ 0.1 mm$^{3}$ MOT volume, which is at the center of a four conducting rod structure which allows us to apply an electric field pulse to ionize the Rydberg atoms and drive the resulting ions to a micro-channel plate detector (MCP). The MOT cooling and repump lasers provide a steady atom population in the $5p_{3/2}$ state.

As shown in Fig. \ref{termdiag}, we excite the atoms from the $5p_{3/2}$ state to the Rb $35f$ and $38p_j$ states by tuning the microwave frequency near the $36d_j \rightarrow 35f$ and $38s_{1/2} \rightarrow 38p_j$ frequencies and the 480 nm beam near the $5p_{3/2} \rightarrow 36d_j$ and $5p_{3/2} \rightarrow 38s_{1/2}$ frequencies. For reference, the microwave frequencies relevant to this work are given in Table \ref{freqs}. To generate the 480 nm pulse the 25 mW output of a 960 nm Toptica single mode diode laser is amplified to 180 mW with a tapered amplifier and further amplified with a two stage pulsed dye amplifier. The dye amplifiers are each pumped with 4 mJ pulses of light at 532 nm from an injection seeded Nd:YAG laser running at a 15 Hz repetition rate. The amplified 960 nm beam is frequency doubled to 480 nm in a potassium niobate crystal. The 480 nm beam can be further amplified using a dye amplifier pumped with $\sim$1 mJ pulses of the 355 nm third harmonic of the Q-switched Nd:YAG laser, resulting in a pulse of 5 ns duration. The spectral line width of the 480 nm light is 150 MHz, the maximum pulse energy is $\sim$10 $\mu$J, and the beam is weakly focused to a diameter of 0.5 mm where it crosses the MOT.

The microwaves are generated by an Agilent E8247C synthesizer, which is operated near 13 GHz for the excitation of the $35f$ state and 19 GHz for the excitation of the $38p_j$ states. The continuous wave output of the synthesizer is formed into pulses using a General Microwave DM862B switch. The pulses are frequency doubled with a Narda DBS-2640X220 active doubler, which produces 26 or 38 GHz pulses with powers of 60 mW. The pulses are doubled again using a Pacific Millimeter V2W0 passive multiplier to produce 52 and 76 GHz pulses with peak powers of up to 10 mW. The microwaves then pass through a Millitech DRA-15-R0000 precision attenuator to a horn outside a window of the vacuum chamber. Although we have no direct way of measuring microwave powers at frequencies above 40 GHz, we can determine the microwave intensity at the MOT by observing the Rabi or Autler Townes splitting when the microwaves are tuned to the $36d\rightarrow35f$ or $38s\rightarrow38p_{1/2}$ resonances \cite{aut55,shir65}.  Comparing the microwave intensities implied by the Rabi frequencies to the estimated power from the multiplier chain indicates that roughly 10\% of the microwave power radiated from the horn reaches the atoms in the MOT.

The timing of the experiment is as follows. The microwave pulse is turned on 1 $\mu$s before the laser pulse and turned off 50 ns after it. A field ionization pulse with a 1.0 $\mu$s rise time (time from 10\% to 90\% of the maximum pulse amplitude) is applied 100 ns after the end of the microwave pulse. The time resolved field ionization signal from the MCP is captured by one or two gated integrators and stored in a computer for later analysis. To develop a sense of the usable range of detunings from the intermediate states we recorded spectra both by scanning the microwave frequency with the laser frequency fixed, and by fixing the microwave frequency and scanning the laser frequency. The relative laser frequency was recorded by monitoring the transmission of the 960 nm light through a 1.5 GHz free spectral range confocal interferometer. The frequencies are measured relative to the known $5p_{3/2} \rightarrow 36d_j$ and $5p_{3/2} \rightarrow 38s$ frequencies \cite{sana15}. Scanning either the microwave or the optical frequency provides equivalent information, and, since they are more intuitive, we present here only scans of the laser frequency at fixed microwave frequency.

\section{Observations}

\begin{table}[]
\caption{Frequencies of the microwave transitions observed in this experiment.}
\begin{center}
\begin{tabular}{|c|c|}
\hline
& \\
\ \ \ Transition\ \ \ &\ \ \ Frequency (GHz)\ \ \ \\
& \\
\hline
& \\
\ \ \ $38s_{1/2} \rightarrow 38p_{1/2} \ \ \ $&72.422 \\
\ \ \ $38s_{1/2} \rightarrow 38p_{3/2} \ \ \ $&74.389 \\
\ \ \ $38s_{1/2} \rightarrow 39s_{1/2}$ \ \ \ &74.386 \\
\ \ \ $36d_{5/2} \rightarrow 35f_{7/2}$ \ \ \ &51.359 \\
\ \ \ $36d_{5/2} \rightarrow 37p_{3/2}$ \ \ \ &47.399 \\
\ \ \ $36d_{3/2} \rightarrow 37p_{1/2}$ \ \ \ &49.285 \\
& \\
\hline
\end{tabular}
\end{center}
\label{freqs}
\end{table}%

\begin{figure}
\centerline{\resizebox{0.50\textwidth}{!}{\includegraphics{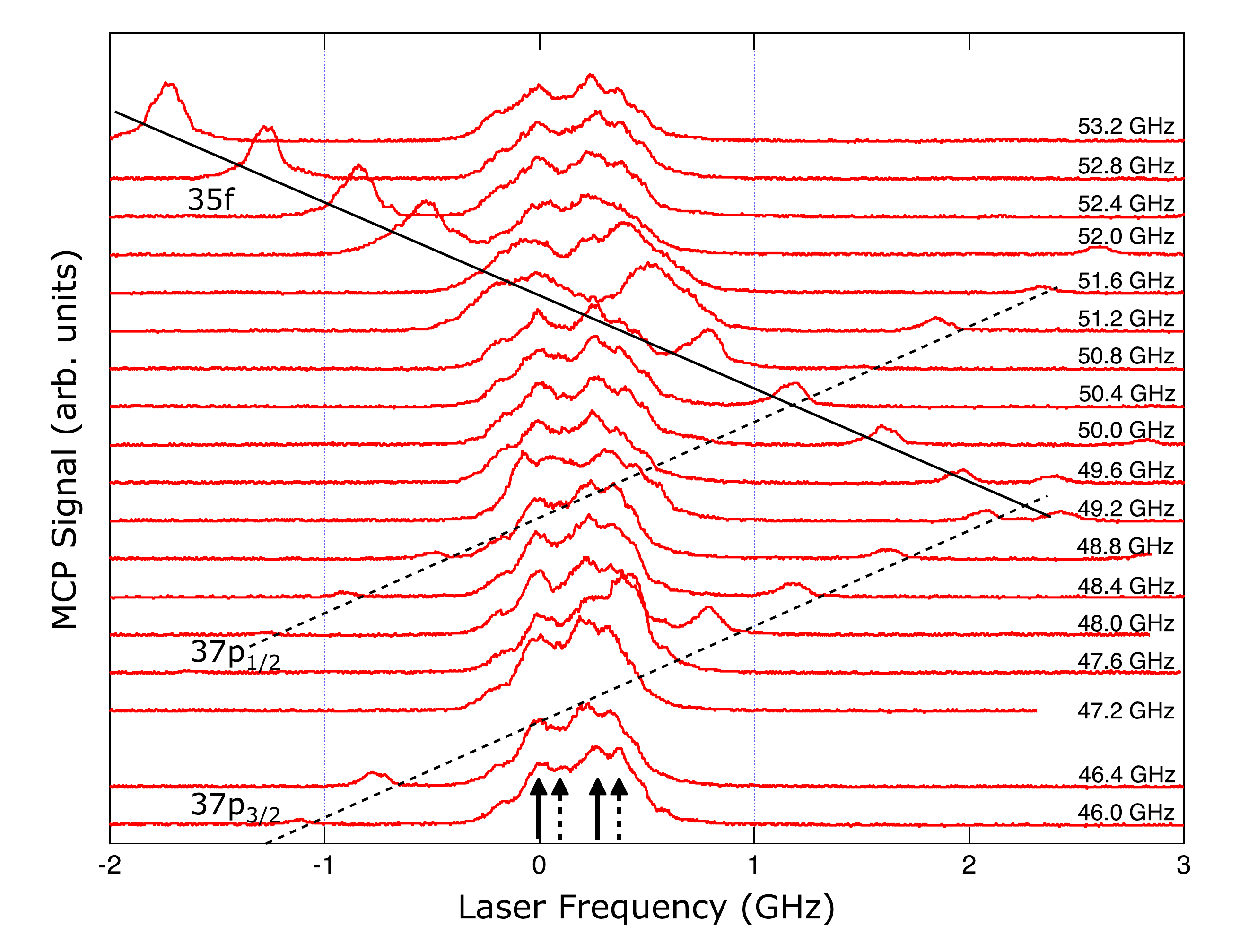}}}
\caption{(color online).
Laser frequency scans of the $5p_{3/2} \rightarrow 37p$ and $\rightarrow 35f$ two photon transitions obtained using fixed microwave frequencies. The 481 nm laser frequencies are relative to the $5p_{3/2} (F=4)\rightarrow 36d_{3/2}$ transition frequency, and the microwave frequency used is noted on each trace. On the spectrum obtained using a microwave frequency of 46.0 GHz (bottom trace), the two solid arrows denote the positions of the $5p_{3/2} (F=4) \rightarrow 36d_{3/2}$ (left) and $5p_{3/2} (F=4) \rightarrow 36d_{5/2}$ (right) transitions, while the dashed arrows denote the positions of the $5p_{3/2} (F=3) \rightarrow 36d_{3/2}$ (left) and $5p_{3/2} (F=3) \rightarrow 36d_{5/2}$ (right) transitions. (The $36d_{5/2} - 36d_{3/2}$ fine structure splitting is 257 MHz, and the $^{85}$Rb $5p_{3/2} (F=4) - (F=3)$ hyperfine interval is 121 MHz.) Visible in the spectra are the $5p_{3/2} \rightarrow 37p_{5/2}$ and $\rightarrow 37p_{3/2}$ two photon transitions (connected by broken lines) and the $5p_{3/2} \rightarrow 35f$ transition (connected by the solid line). For reference, the $37p$ fine structure splitting is 2.144 GHz}
\label{36dspectra}
\end{figure}

Fig. \ref{36dspectra} shows the two photon excitation spectra from the $5p_{3/2}$ state to the $35f$ and $37p_j$ states, using the $36d$ state as the near resonant intermediate state. These spectra were obtained by sweeping the laser frequency at fixed microwave frequency, and we obtained the spectra for a range of microwave frequencies. For reference, the resonant frequencies of the relevant microwave transitions are shown in Table \ref{freqs}. The fine structure splitting of the $35f$ state, 5 MHz, is not resolved. The $36d$ fine structure splitting (257 MHz) should be resolvable with our pulsed laser system, but since the MOT cooling and repump lasers populate both $F=3$ and $F=4$ hyperfine states of $5p_{3/2}$ (which are separated by 121 MHz), and these both couple to both the $36d_{3/2}$ and $36d_{5/2}$ states, we also see features with spacings of 121 MHz. This is described more fully below, and the net effect is that the central feature, due to the laser being resonant with the $5p_{3/2} \rightarrow 36d$ manifold, is poorly resolved.

The zero of the frequency scale of Fig. \ref{36dspectra} is the frequency of the $5p_{3/2} (F=4) \rightarrow 36d_{3/2}$ transition. The 480 nm power is adequate to slightly saturate the $5p_{3/2}\rightarrow 36d$ transition, and the estimated microwave intensity at the atoms is 0.3 mW/cm$^2$ (4 mW emitted from the horn). With our field pulse, the field ionization signals of the Rb $36d$, $35f$, and $37p$ states are detected at almost identical times. The Fig. \ref{36dspectra} spectra therefore represent the entire Rydberg excitation spectra.

It is useful to first consider a scan in which the microwave frequency is far off resonant, for example at 46.0 GHz. At this frequency we see several peaks in the spectrum around the $5p_{3/2}\rightarrow 36d_j$ frequencies. The solid arrows show the transitions from the $5p_{3/2}$ $F=4$ state to the $36d_{3/2}$ state (near 0 GHz) and to the $36d_{5/2}$ state (near 0.3 GHz), which are separated by 257 MHz. Displaced by 121 MHz to higher laser frequency are the transitions from the $5p_{3/2}$ $F=3$ state to the $36d_j$ states, marked with broken arrows. From the 46.0 GHz trace of Fig. \ref{36dspectra} these assignments are not so obvious, but at lower 480 nm power the subsidiary peaks at relative frequencies of -200 MHz and 500 MHz disappear. The microwave frequency 51.2 GHz is approximately resonant with the $36d\rightarrow35f$ transition, resulting in a 500 MHz Rabi splitting of the two main peaks, from which we infer a microwave field amplitude and intensity of 0.5 V/cm and 0.3 mW/cm$^2$, respectively \cite{walker08, edmo74}. At microwave frequencies of 52.0 GHz and higher the two photon microwave-optical excitation of the $35f$ state is visible at laser frequencies below the $5p_{3/2}\rightarrow 36d$ frequency. At a microwave frequencies of 50.8 GHz and lower the two photon transition is seen at laser frequencies above the $5p_{3/2}\rightarrow 36d$ frequency. In all cases the two photon $5p_{3/2}\rightarrow 35f$ resonances lie on the calculated location of the $5p_{3/2}\rightarrow 35f$ resonance shown by the solid line. As shown, the magnitude of the $35f$ signal is comparable to that of the $36d$ signal. Using the $36d$ state as the intermediate state, two photon excitation of the $37p_j$ states is also possible, and these excitations can be seen to fall on the broken lines of Fig. \ref{36dspectra}, their calculated locations.

\begin{figure}
\centerline{\resizebox{0.50\textwidth}{!}{\includegraphics{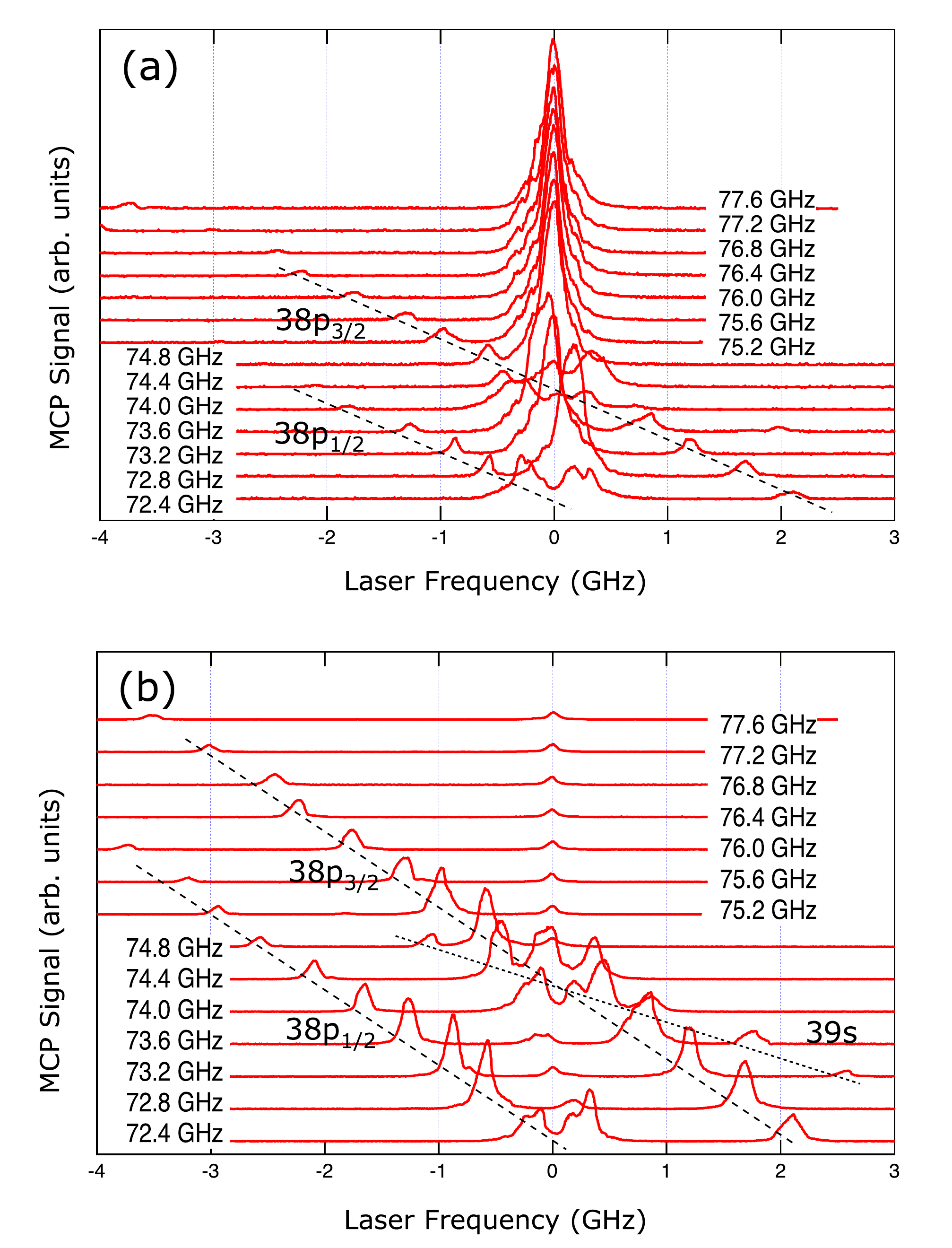}}}
\caption{(color online).
Laser frequency scans of the $5p_{3/2} \rightarrow 38p$ two photon transitions obtained using fixed microwave frequencies. The 481 nm laser frequencies are relative to the $5p_{3/2} (F=4)\rightarrow 38s_{1/2}$ transition frequency, and the microwave frequency used is noted on each trace. In (a) we show the spectra obtained when the gated integrator is positioned on the $38s$ field ionization signal, while in (b), the spectra obtained with the gate positioned on the $38p$ field ionization signal are shown. Visible in the spectra are the $5p_{3/2} \rightarrow 38p_{3/2}$ and $\rightarrow 38p_{1/2}$ two-photon transitions (connected by broken lines) and the $5p_{3/2} \rightarrow 39s$ three-photon transition (connected by the dotted line). For reference, the $38p$ fine structure splitting is 1.968 GHz}
\label{38sspectra}
\end{figure}

The spectra of Fig. \ref{36dspectra} show that we are able to resolve spectrally the $36d$, $35f$, and $37p$ states, but they do not demonstrate that the excitation is selective, that is, that when we tune to the two photon $5p_{3/2}\rightarrow35f$ resonance we only produce $35f$ atoms, not $36d$ atoms. To show that the two photon excitation results in only the desired state we have examined the excitation of the $38p_j$ states using the $38s$ state as the near resonant intermediate state, as shown in Fig. \ref{termdiag}. Since the $n=35$  $\ell\geq 3$ states lie energetically between the $38s$ and the $38p_j$ states, the $38p_j$ states require a 10\% lower field for ionization than does the $38s$ state and are ionized earlier in the rising field pulse \cite{gall94,gall77}. As a result, the $38s$ and $38p_j$ state field ionization signals are almost completely resolved in time. In Fig. \ref{38sspectra} we show the spectra observed by setting separate gates on the $38s$ and $38p_j$ field ionization signals and scanning the laser frequency for fixed microwave frequencies between 72.4 and 77.6 GHz.

In Fig. \ref{38sspectra}(a) we show the spectra obtained from the gate on the $38s$ signal. For most of the microwave frequencies shown in Fig. \ref{38sspectra}(a) we observe a strong signal at the frequency of the $5p_{3/2}\rightarrow38s$ transition and almost nothing else. Small signals are seen along the dashed lines, where the excitation of the $38p_{1/2}$ and $38p_{3/2}$ states is calculated to occur, and there is evident 500 MHz Rabi splitting at the $38s\rightarrow38p_{1/2}$ microwave resonance at 72.4 GHz, from which we infer a microwave field amplitude of 0.8 V/cm \cite{walker08, edmo74}. Near the $38s\rightarrow38p_{3/2}$ resonance at 74.4 GHz there is a triple peak, due to the coincidence with the $38s\rightarrow39s$ two photon resonance, as shown by Table \ref{freqs}.

With the gate set on the $38p$ field ionization signal we observe the spectra shown in Fig. \ref{38sspectra}(b). We see almost no signal at the laser frequency of the $5p_{3/2}\rightarrow38s$ transition, and strong signals are observed along the dashed lines, demonstrating that in the two photon excitations we are only exciting the $38p_{1/2}$ and $38p_{3/2}$ states, not the $38s$ state. At 73.6 GHz and 74.8 GHz additional well resolved peaks are observed, corresponding to the three photon excitation of the $39s$ state, which is calculated to fall on the dotted line. Thus, while the data of Fig. \ref{36dspectra} do not explicitly show that at the two photon $5p_{3/2}\rightarrow 35f$ resonances we are exciting only the $35f$ state, but not the $36d$ state, the data shown in Fig. \ref{38sspectra}(b) certainly support that conclusion, since we see negligible excitation of the $38s$ intermediate state at the $5p_{3/2}\rightarrow 38p_j$ two photon resonances.

\begin{figure}
\centerline{\resizebox{0.50\textwidth}{!}{\includegraphics{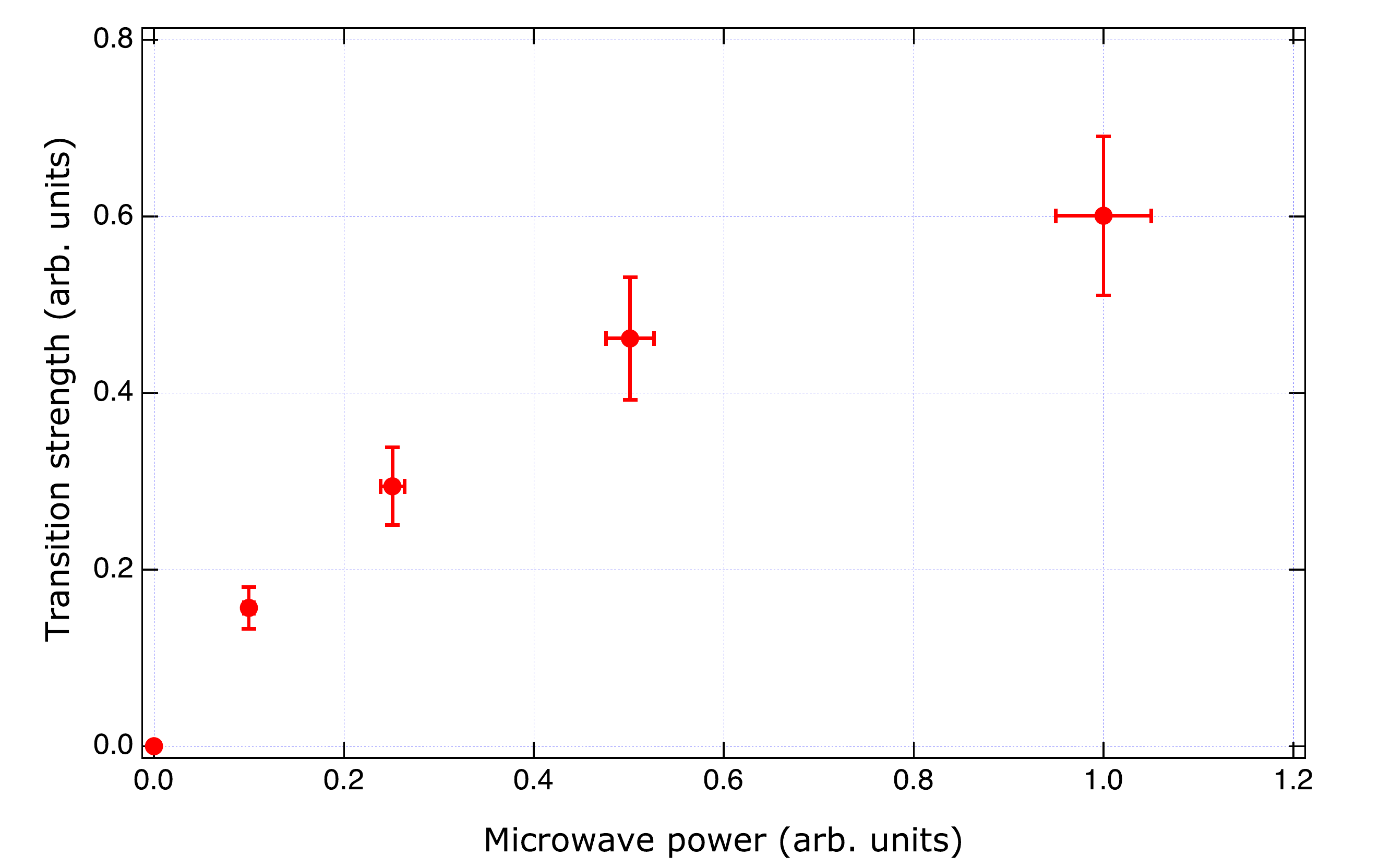}}}
\caption{(color online).
Strength of the $5p_{3/2} \rightarrow 35f$ two photon transition as a function of microwave power at a microwave frequency of 53.4 GHz. This is the area of the $5p_{3/2} \rightarrow 36d \rightarrow 35f$ peak (i.e., the leftmost peak in the top trace of Fig. \ref{36dspectra}), which is normalized to the area of the on-resonance $5p_{3/2} (F=4) \rightarrow 36d_{3/2}$ transition (i.e., the peak at 0 GHz). A relative microwave power of 1.0 corresponds to the emission of 4 mW from the microwave horn. As expected, at low power the $35f$ signal increases linearly with microwave power, but at a relative power of 0.5, it begins to saturate.}
\label{Graph4}
\end{figure}

As shown by Fig. \ref{36dspectra}, the two photon excitation of the $35f$ state is comparable to the excitation of the $36d$ state. A different measure of the efficiency of the two photon excitation of the $35f$ state is provided by the microwave power dependence shown in Fig. \ref{Graph4}. The microwave frequency is 53.2 GHz, which is detuned from the $36d_{5/2}\rightarrow 35f$ resonance frequency by 1.8 GHz. The 480 nm power is relatively high, so that the transition to the $36d$ state is saturated. At low power the $35f$ signal increases linearly with microwave power, but at a relative power of 0.5 in Fig. \ref{Graph4}, it begins to saturate.

\section{Summary and outlook}

We have described an efficient microwave-optical two photon excitation of cold Rydberg atoms which allows access to states not normally accessible optically. The method uses modest laser and microwave sources and is made possible by the combination of cold atoms, a relatively narrow band pulsed laser, and the large electric dipole matrix elements of the Rydberg atoms. The limitation in the present experiments is the 150 MHz bandwidth of the 480 nm light, but the intrinsic limitation for excitation from the Rb $5p_{3/2}$ state is its radiative width of 6 MHz. Using laser pulses of comparable linewidth to the $5p_{3/2}$ state natural width, and correspondingly longer duration than those we used, would allow the detunings to be reduced by an additional factor of ten, with corresponding decreases in the requisite optical and microwave powers. With such laser pulses excitations involving the absorption of more microwave photons and adiabatic rapid passage begin to look attractive. Finally, using excitation from the ground state removes the limitation imposed by rapid decay of the $5p_{3/2}$ state. Detunings of only a few MHz can be used, further reducing the requisite optical and microwave powers.

\section {Acknowledgements}
It is a pleasure to acknowledge useful conversations with R. Ali, R. R. Jones, W. Li, and T. Vogt, and C. Gross. This work has been supported by the Air Force Office of Scientific Research under grant FA9550-14-1-0288 and by Colby College.


%

%

\end{document}